# Electronic states and molecular orientation of ITIC film[*]

Du Ying-Ying(杜莹莹)[a], Lin De-Qu(林德渠)[a], Chen Guang-Hua(陈光华)[a], Bai Xin-Yuan(白新源)[a], Wang Long-Xi(汪隆喜)[a], Wu Rui(吴蕊)[b], Wang Jia-Ou(王嘉鸥)[b], Qian Hai-Jie(钱海杰)[b], and Li Hong-Nian(李宏年)[a,†]

[a] *Department of Physics*, *Zhejiang University*, *Hangzhou* 310027, *China*

[b] *Institute of High Energy Physics*, *Chinese Academy of Sciences*, *Beijing* 100049, *China*

ITIC is the milestone of non-fullerene small molecule acceptors used in organic solar cells. We have studied the electronic states and molecular orientation of ITIC film using photoelectron spectroscopy and X-ray absorption spectroscopy. The negative integer charge transfer energy level is determined to be 4.00±0.05 eV below the vacuum level, and the ionization potential is 5.75±0.10 eV. The molecules predominantly adopt the face-on orientation on inert substrates as long as the surfaces of the substrates are not too rough. These results provide physical understanding of the high performance of ITIC-based solar cells, also afford implications to design more advanced photovoltaic small molecules.

Keywords: Electronic states, molecular orientation, photovoltaic small molecule, photoelectron spectroscopy, XAS

PACS: 88.40.jr, 73.30.+y, 61.46.-w, 61.05.cj

[*]Project supported by the National Natural Science Foundation of China (Grant Nos. 11374258 and 11079028).
[†]Corresponding author. E-mail: phylihn@mail.zju.edu.cn

## 1. Introduction

Organic solar cells (OSCs) have attracted extensive attentions during the past two decades due to the advantages of light weight, mechanical flexibility, low cost, and large-scale solution-fabrication. This type of solar cell comprises a thin layer of electron donor/electron acceptor blend sandwiched between two electrodes. Fullerene derivatives such as $PC_{61}BM$[1] and $PC_{71}BM$[2] have been the most popular electron acceptors. However, fullerene derivatives have the drawback of very weak photo-absorption in the visible and near-infrared regions. Therefore, high-performance non-fullerene acceptor is highly desirable. In the very recent years a few non-fullerene acceptors have been developed, which exhibited comparable photovoltaic property with $PC_{61}BM$ or $PC_{71}BM$. Among them ITIC is the milestone molecule.[3] The OSC based on ITIC exhibited power conversion efficiency of 11.21%,[4] and the existing more advanced non-fullerene acceptors are the modification of ITIC.[5-11]

The energy position of the highest-occupied-molecular-orbital (HOMO) and the lowest-unoccupied-molecular-orbital (LUMO) plays crucial roles in the performance of OSCs. The HOMO level, or the top of the HOMO band ($E_{HOMO}(top)$) for solid samples, is the minus of ionization potential (IP) if referenced to the vacuum level. The IP of ITIC was measured to be 5.48-5.67 eV with the cyclic voltammetry method for solution samples.[4,11-16] Considering the fact that ITIC behaves as solid state in OSCs, the IP of a film sample is desirable. One purpose of our work is to obtain the IP of ITIC film with ultraviolet photoelectron spectroscopy (UPS) measurements.

For electron acceptor materials such as ITIC the LUMO level is more pertinent to the performance of OSCs as compared with the HOMO level. The energy position of the bottom of the LUMO band ($E_{LUMO}(bottom)$) is the minus of electron affinity (EA). The EA was measured to be 3.63-4.00 eV with the cyclic voltammetry method for solution samples.[4,11-16] The result for film sample has not been reported in the literature. Actually, the concept of the LUMO level (or $E_{LUMO}(bottom)$) should be substituted by the concept of the negative integer charge transfer level ($E_{ict-}$) in many aspects of the performance of OSCs.[17] The energy position of $E_{ict-}$ is slightly lower than the $E_{LUMO}$(bottom), i.e., the absolute value of $E_{ict-}$ is slightly greater than EA. More detail about the concept of $E_{ict-}$ can be found in Refs. [18−21] and will be narrated later. $E_{ict-}$ can be measured with UPS, and the measured $E_{ict-}$ has been reported for many organic molecules.[17,19,22,23] Although there are a few UPS data of ITIC film in Refs. [7] and [16], the $E_{ict-}$ has not been determined. The main purpose of our work is measuring the $E_{ict-}$ of ITIC film.

Molecular orientation has significant effect on charge transport. For planar organic molecules face-on (with the molecular plane parallel to the substrate) and edge-on (with the molecular plane perpendicular to the substrate) are two frequently encountered orientations. The face-on orientation is more beneficial to facilitate the charge transport in OSCs, since the stacking of the π (and π*) orbital forms the charge transport channel across the device.[24-26] The molecular orientation in ITIC film was found to be substrate-dependent according to some grazing incidence wide-angle X-ray scattering (GIWXAS) measurements.[13,14,27,28] To get a deep understanding of the molecular orientation, we prepared ITIC film samples on some different inert substrates and performed X-ray absorption spectroscopy (XAS) measurements. The XAS data also provide some information about the unoccupied electronic states.

## 2. Experiment

We utilized three substrates, naturally oxidized Al(111) single crystal, ITO glass, and Si:H(111), to study the electronic states and the molecular orientation. The Al(111) single crystal was simply cleaned by sonication in de-ionized water and acetone. The ITO glass was ultrasonically cleaned with detergent for 30 min, washed with much de-ionized water, and further ultrasonically cleaned with de-ionized water, acetone, and isopropyl alcohol. The Si:H(111) substrate was prepared as follows. Si(111) wafers were first ultrasonically cleaned in deionized water and ethyl alcohol. Then the wafers were rinsed sequentially in solution 1 ($NH_3·H_2O:H_2O_2:H_2O$=1:1:6, 80 °C, 15 min) and solution 2 ($HCl:H_2O_2:H_2O$=1:1:6, 80 °C, 15 min). After being washed in deionized water, the wafers were etched in 5% HF for 1 min to obtain the H-passivated Si(111) surface. At last, the Si:H(111) wafers were again rinsed in deionized water for less than 5 s to remove residual compounds from the etching process. After cleaning, the substrates were dried by nitrogen flow and transferred into a nitrogen-filled glove box immediately. In the following we abbreviate the naturally oxidized Al(111) single crystal, the ITO glass, and the Si:H(111) to be Al, ITO, and Si, respectively.

ITIC powder (purity > 99%) was purchased from the Solenne company. The powder was dissolved in chlorobenzene at a concentration of 25 mg/mL. The solution was stirred at 70 °C for ~24 hours and filtrated through a polytetrafluoroethylene filter (0.2 μm). ITIC films were then spin coated (2000 rpm for 60 s) on the substrates. After the spin coating we annealed the samples on a hot plate at 130 °C for 10 min. All the above experimental operations were carried out in the glove box. The thickness of the films was not quantitatively calibrated but was proper for our study. The signal of the substrates were totally attenuated, and charging effect was not observed in the UPS measurements.

The ITIC film samples were placed in a small plastic box in the glove box, and the plastic box was well sealed with parafilm before fetched out. Spectral measurements were carried out at the Photoelectron Spectroscopy Endstation of the Beijing Synchrotron Radiation Facility (BSRF). The samples were clamped onto stainless steel sample holders in a helium-filled glove box of BSRF and then transported into the vacuum system of the endstation; the time exposing to air was less than 30 s in this process. We measured the spectra after the samples were left in the ultra high vacuum environment ($2\times10^{-10}$ mbar) for more than six hours. The UPS spectra were measured with the photon energy of 21.2 eV and a VG Scienta R4000 analyzer. The sample normal coincided with the entrance of the energy analyzer, and the photoelectrons with different emission angles (within ±19° with respect to the sample normal) were integrated. In work function measurements we applied –5.0 V bias to the samples. The overall energy resolution was better than 0.05 eV. X-ray photoelectron spectra (XPS) were also measured for a portion of the samples with the photon energy of 700 eV; the overall energy resolution was ~0.75 eV. Carbon K-edge XAS spectra were acquired with the total electron yield mode; the energy resolution was ~0.2 eV. For helping to understand the XAS data we slightly doped the ITIC/Si sample by depositing Ca atoms onto the sample surface. The Ca atoms were evaporated from a Ta boat mounted in the ultra high vacuum system.

C and O contamination should be well controlled for reliable spectra measured on ex-situ prepared samples. ITIC itself contains both C and O atoms, and the possible contamination cannot be checked by UPS or XPS measurements. So we prepared and measured some $PC_{61}BM$ and P3HT samples (the latter molecule does not contains O atom) to check the contamination. The measured UPS spectrum of the spin coated $PC_{61}BM$ film was unexpectedly the same as that of the $PC_{61}BM$ film prepared by vacuum deposition.[29] The O 1s signal could not be observed in the XPS spectrum of the P3HT sample. So the ITIC samples of this work were very clean, and the measured spectra are of high quality.

For helping to understand the experimental data we performed density functional theory (DFT) calculations for isolated ITIC molecule. The calculating details are the same as that we described previously.[29,30]

## 3. Results and discussion

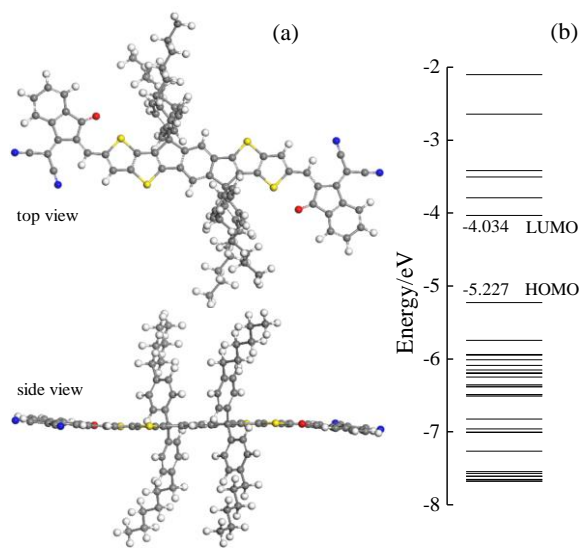

**Fig. 1.** (color online) (a) Optimized structure of ITIC by DFT calculations. The gray, white, red, blue, and yellow balls represent C, H, O, N, and S atoms, respectively. (b) Energy level diagram near the HOMO and LUMO levels.

Figure 1 shows the optimized structure of ITIC molecule and the energy level diagram near the HOMO and LUMO levels. The main chain where most of the π electrons are located is planar. The energy levels are for ground state, and the HOMO-LUMO gap (1.193 eV) in Fig. 1(b) should be much smaller than experimentally measured gap. However, the energy intervals between the occupied levels are generally (though not always) reliable.[29-31]

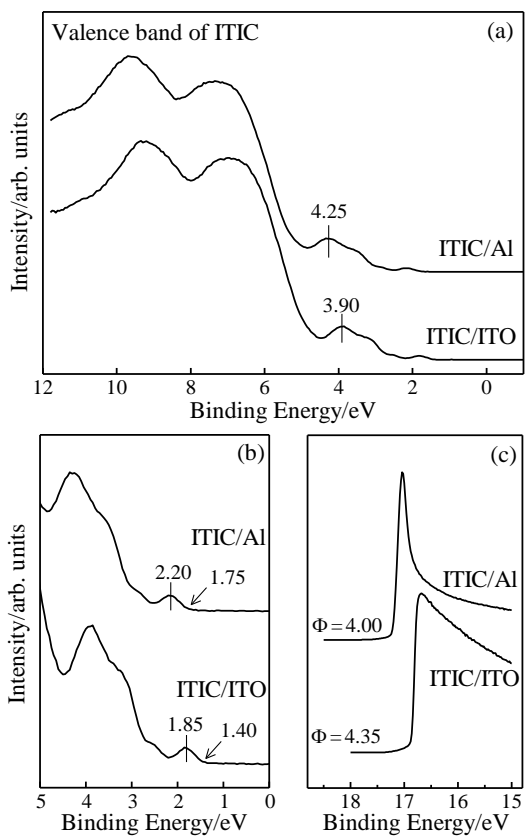

**Fig. 2.** (a) UPS data of ITIC films on Al and ITO substrates. (b) Enlarged view of the photoemission between the Fermi level (0 eV) and 5.0 eV binding energy. (c) Secondary electron cutoffs. The work function data indicated in (c) are determined by the tangent of the cutoff and the base line.

Figure 2(a) shows the UPS data of the ITIC/Al and ITIC/ITO samples. The spectra in Fig. 2(a) are normalized to the height of the feature at 4.25 eV in the ITIC/Al spectrum (3.90 eV in the ITIC/ITO spectrum). The reason for the different binding energies of the two spectra will be explained later. Figure 2(b) enlarges the spectral features between the Fermi level and 5 eV binding energy; the spectral shape is much alike the reported spectra.[7,16] Comparing Fig. 2(b) with Fig. 1(b), the calculated energy levels correctly predict that the HOMO signal is well separated from the signal of the other occupied levels. The center of the HOMO photoemission locates at 2.20 eV for the ITIC/Al sample and 1.85 eV for the ITIC/ITO sample. We do not show the UPS spectra of the ITIC/Si sample because they cannot provide more information than what can be drawn from Fig. 2.

UPS is the most powerful tool to measure the IP of a solid sample. IP is the energy difference between the $E_{HOMO}$(top) and the vacuum level. According to Fig. 2(b) the $E_{HOMO}$(top) locates at ~1.75 and ~1.40 eV for the ITIC/Al and ITIC/ITO samples, respectively. The vacuum level (identical to the work function Φ) is indicated in Fig. 2(c). The value of the IP obtained for the ITIC/Al sample (1.75+4.00 = 5.75 eV) is the same as that for the ITIC/ITO sample (1.40+4.35 = 5.75 eV), which is reasonable because IP has little relation with the substrates. The experimental error, including some arbitrariness in deciding the energy position of the $E_{HOMO}$(top) in Fig. 2(b), is at most ±0.10 eV. The IP of 5.75±0.10 eV for our film samples is some different from that (5.48-5.67 eV)[4,11-16] measured by cyclic voltammetry. Our result is more practical since ITIC behaves as solid state in OSCs.

UPS is also the most powerful tool to measure $E_{ict-}$. In this case the substrate has to be properly selected. For understanding this we explain the concept of $E_{ict-}$ in some detail. $E_{ict-}$ was proposed for the energy level alignment at the interface between an electron acceptor material and an inert substrate.[18-21] This kind of interface is frequently encountered in OSCs fabricated by spin coating. Overlapping of the electronic wave functions is negligible at the interface, but charge transfer across the interface can be realized by tunneling of integer charge. If the work function of the substrate is sufficiently small, some electrons can transfer from the substrate to the $E_{ict-}$ of the acceptor material. $E_{ict-}$ is lower in energy than the $E_{LUMO}$(bottom) because of two reasons. First, the state accommodating the electrons transferred from the substrate may be the negative polaron (or bi-polaron) state due to the soft nature of many organic molecules.[32] It is well known that the negative polaron level is lower than the $E_{LUMO}$(bottom). Second, the disordered structure of an organic film can induce tail state of the LUMO band extending into the HOMO-LUMO gap by several tenth of eV.[20,33,34] The tail state can also accommodate the electrons transferred from the substrate to pin the Fermi level although the density of states (DOS) of the tail state is generally too weak to be detected by conventional UPS measurements.

According to the meaning of $E_{ict-}$, the substrate for measuring the $E_{ict-}$ of an organic acceptor material must be inert and have small work function. The Al (with natural oxide layer) and ITO substrate are both inert. The work function of our Al substrate is 3.45−3.53 eV as measured on several samples (without the ITIC overlayer); the work function of the ITO substrate is 4.30−4.50 eV. In Fig. 2(c) the work function of the ITIC/Al sample (4.00 eV) is greater than that of the substrate, indicating the electron transfer from the Al substrate to the ITIC film. The Fermi level is then pinned at the $E_{ict-}$, and the measured work function is identical to the $E_{ict-}$ (the value of $E_{ict-}$, by convention, is the absolute value of the energy position of $E_{ict-}$ referenced to the vacuum level). Considering the experimental error, we obtain the conclusion of $E_{ict-}$ = 4.00±0.05 eV for ITIC. The $E_{ict-}$ of ITIC is near that of $PC_{61}BM$ (~3.94 eV)[17], which is one of the reasons why ITIC can substitute for fullerene acceptors.

As comparison, the work function of the ITIC/ITO sample is 4.35 eV, near that of the ITO substrate. This is because that the work function of the ITO substrate (4.30-4.50 eV) is greater than the $E_{ict-}$ (4.00 eV). The electron cannot transfer from the ITO substrate to the ITIC film. The electrons cannot yet transfer from the ITIC film to the ITO substrate, since the LUMO level is empty and the HOMO level is much lower than the Fermi level of the substrate (see Fig. 1(b)). The ITIC/ITO interface is thus vacuum level alignment, and the work function of the ITIC/ITO sample should be the same as that of the substrate. Owing to the different scheme of energy level alignment, the valence band features also present different binding energies for the ITIC/Al and ITIC/ITO samples in Fig. 2(a) and (b).

The Carbon K-edge XAS spectra of ITIC/Al, ITIC/ITO, and ITIC/Si are shown in Fig. 3. The spectral shape, peak position, and angle dependence are much similar for the three samples. The absorption from the C 1s states to the localized π* or σ* orbitals distributes between 284.0−290.5 eV; above 290.5 eV is the absorption to the continuum states (ionization). According to the $E_{ict-}$ of 4.00 eV the distribution of the localized π* and σ* orbitals should be within an energy range of 4.00 eV. The seeming discrepancy is due to the broad distribution of the C 1s states (the initial states of the photon absorption) as indicated in the inset of Fig. 3(a).

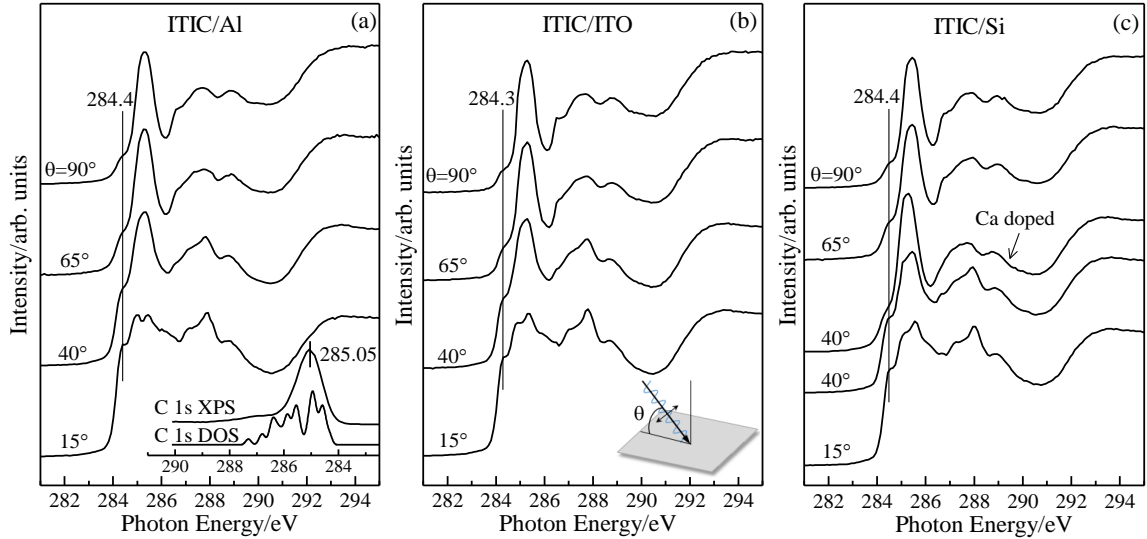

**Fig. 3.** Angle-dependent C K-edge XAS spectra of ITIC film on the Al (a), ITO (b), and Si (c) substrates; the spectra are normalized to the intensity at $h\upsilon$ = 295 eV in each panel. In (c) the spectrum of the Ca-doped ITIC/Si (indicated by an arrow) is also shown for comparison. The inset in (a) is the C 1s XPS spectrum of the ITIC/Al and the C 1s DOS calculated for ITIC molecule; the energy of the DOS has been shifted to coincide with the binding energy of the XPS spectrum. The inset in (b) depicts the definition of θ.

The first feature of the XAS spectra locates at 284.4 eV in Fig. 3(a) and (c) (or 284.3 eV in Fig. 3(b), the difference is within the experimental error). The first feature generally corresponds to the absorption to the LUMO (and possibly the adjacent LUMO+1). Considering the broad distribution of the C 1s states, however, the final states of the first feature should be checked for the samples studied here. The spectrum of the Ca-doped ITIC/Si affords the necessary verification. This spectrum was measured at θ = 40° and is shown next to the 40° spectrum of the un-doped sample in Fig. 3(c). The intensity of first feature is obviously decreased for the doped sample as compared with that of the un-doped sample. We did not perform the doping experiments in detail, and the sample was only slightly doped. Only the LUMO or at most the LUMO+1 were occupied by the electrons transferred from the Ca atoms. So the decreased intensity of the first feature reveals that the final states is mainly composed of the LUMO and LUMO+1 states. This conclusion is the premise of studying the molecular orientation.

In XAS measurement the absorption cross section of a K-shell (here C 1s) resonance depends on the projection of the X-ray electric field vector onto the final state orbitals involved in the transition. The absorption intensity of a C 1s−>π* transition is the maximum/zero if the electric field vector is parallel/perpendicular to the π* orbital. The main chain of ITIC is planar (Fig. 1(a)); the LUMO and LUMO+1 are π* nature according to our DFT calculations. So the molecular orientation in the ITIC film can be deduced from the angle-dependent XAS spectra. In Fig. 3 θ means the angle between the incoming X-ray beam and the sample surface, as indicated by the inset in Fig. 3(b). The first feature at 284.4 (or 284.3) eV mainly corresponds to the C 1s−>LUMO (probably also LUMO+1) transition as discussed previously. The intensity of this feature changes remarkable with different angles. So the ITIC molecules in the film have significant preferable orientation. More specifically, the intensity of the first feature decreases monotonically with increasing the angle from 15° to 90°. Considering the fact that the LUMO and LUMO+1 orbitals (π*) are perpendicular to the plane of the main chain of ITIC, the molecules predominantly adopt the face-on orientation on all the three substrate.

The substrates used in this work and in actual OSC fabrication are all inert, and thus the interfacial chemical interaction is negligible. What determines the molecular orientation should be the intrinsic inter-molecular interaction and the substrate morphology. The Si:H(111) substrate is atomic smoothing. The naturally oxidized Al(111) substrate is near atomic smoothing. The ITO substrate has the surface roughness of several nm (revealed by atomic force microscopy images not shown here), but the ITIC molecules still predominantly adopt the face-on orientation. Therefore, we think that ITIC molecules prefer to form face-on oriented film on inert substrates as long as the surfaces of the substrates are not too rough. The different degree of the face-on orientation on some other substrates reported in Refs. [13,14,28] should be due to the different surface roughness of the substrates. The highly disordered orientation for the ITIC film on the PEDOT:PSS covered Si wafer[27] was most possibly due to the very rough and mutual penetrating ITIC/PEDOT:PSS interface (the surface layers of the PEDOT:PSS film were re-dissolved when spin coating ITIC).

## 4. Conclusions

The $E_{ict-}$ of ITIC is 4.00±0.05 eV and near that of $PC_{61}BM$, which provides an reference not only for selecting matching donor materials in fabricating OSCs but also for modifying the molecular structure of ITIC to obtain more advanced acceptor materials. ITIC molecules prefer to form face-on oriented film as long as the substrate surface is not too rough. So one can control the charge transporting property of ITIC-based devices by fabricating ITIC/donor and ITIC/electrode interfaces with proper morphology.